\documentclass[preprint,showpacs,preprintnumbers]{revtex4}
\usepackage{graphicx}
\usepackage{dcolumn}
\usepackage{bm}

\begin{document}

\preprint{APS/123-QED}
\title{Comment on "Large Difference in the Elastic Properties of fcc and hcp
Hard-Sphere Crystals"}
\author{Nazar Sushko}
\affiliation{Laboratory of Polymer Chemistry, University of Groningen, Nijenborgh 4, 9747
AG Groningen, The Netherlands }
\author{Paul van der Schoot}
\affiliation{Eindhoven Polymer Laboratories, Eindhoven University of Technology, P.O. Box
513, 5600 MB Eindhoven, The Netherlands}
\date{\today}

\begin{abstract}
As is well known, hard-sphere crystals of the fcc and hcp type differ very
little in their thermodynamic properties. Nonetheless, recent computer
simulations by Pronk and Frenkel indicate that the elastic response to
mechanical deformation of the two types of crystal should be quite different 
\cite{Pronk}. By invoking a geometrical argument put forward by R. Martin
some time ago \cite{Martin}, we suggest that this is largely due to the
different symmetries of the fcc and hcp crystal structures. Indeed, we find
that elastic constants obtained by means of computer simulations for the fcc
hard-sphere crystal can be mapped onto the equivalent ones of the hcp
crystal to very high accuracy. The same procedure applied to density
functional theoretical predictions for the elastic properties of the fcc
hard-sphere crystal also produces remarkably accurate predictions for those
of the hcp hard-sphere crystal.
\end{abstract}

\pacs{62.20.Dc, 82.70.Dd}
\maketitle

In a recent publication \cite{Pronk}, Pronk and Frenkel reported on a
computer simulation study of the elastic properties of fcc and hcp crystals
of hard spheres. They found the various elastic constants to differ by up to
twenty per cent, despite that the thermodynamic properties of both types of
hard-sphere crystal are almost indistinguishable. Indeed, the free energy,
pressure and compressibility of the two crystal types deviate from each
other by less than 0.1\%, at least for conditions not too far from the
melting point \cite{Pronk}. In this Comment, we point out that the
difference in the elasticity of fcc and hcp hard-sphere crystals is less
surprising than claimed by Pronk and Frenkel, and that it can be explained
by the geometry of the packing of the particles within each lattice.

The relation between the elastic moduli of hcp and fcc crystals have been
studied theoretically and experimentally by number of authors. See, e.g., 
\cite{Novotny,Leamy,Martin,Ahmad,Robinson}. Of particular interest is the
work of R. Martin \cite{Martin}, who derived an approximate transformation
of the elastic moduli of the fcc crystal to those of the hcp lattice, making
use of the fact that both lattice types can be constructed from tetrahedral
units. The tetrahedral blocks in the fcc lattice are equally oriented, while
the hcp lattice can be built up from two tetrahedra, oriented differently
from each other and from those of the fcc lattice. The transformation of any
tensor in the fcc system of coordinates (defined as usual along the cubic
axes) to either of the two representations of this tensor in the trigonal
geometry of the hcp crystal can be made by two simple rotations $\boldsymbol{%
R}^{(1)}$ and $\boldsymbol{R}^{(2)}$, where 
\begin{equation}
\boldsymbol{R}^{(1)}=\left( 
\begin{array}{ccc}
\sqrt{\frac{1}{2}} & 0 & \sqrt{\frac{1}{2}} \\ 
-\sqrt{\frac{1}{6}} & \sqrt{\frac{2}{3}} & \sqrt{\frac{1}{6}} \\ 
-\sqrt{\frac{1}{3}} & -\sqrt{\frac{1}{3}} & \sqrt{\frac{1}{3}}%
\end{array}%
\right) \;,\;\boldsymbol{R}^{(2)}=\left( 
\begin{array}{ccc}
-1 & 0 & 0 \\ 
0 & -1 & 0 \\ 
0 & 0 & 1%
\end{array}%
\right) \boldsymbol{R}^{(1)}\;.  \label{transform}
\end{equation}%
This suggests that the transformation of the elastic moduli tensor in the
fcc geometry, $\boldsymbol{C}^{FCC}$, to that of the trigonal geometry of
the hcp lattice, $\overline{\boldsymbol{C}}^{HCP}$, could simply be the
average (superposition) of the two trigonal tensors \cite{Robinson}, 
\begin{equation}
\overline{C}_{ijkl}^{HCP}=\frac{1}{2}\left(
R_{ir}^{(1)}R_{js}^{(1)}R_{kt}^{(1)}R_{lu}^{(1)}C_{rstu}^{FCC}+R_{ir}^{(2)}R_{js}^{(2)}R_{kt}^{(2)}R_{lu}^{(2)}C_{rstu}^{FCC}\right) \;,
\label{aver}
\end{equation}%
where the subscripts have their usual meaning. It so happens, however, that
the two unequal tetrahedra of the hcp lattice are not independent, but
attached to each other. Hence, the elastic response of the hcp lattice to an
external strain should be the \textit{combined} response of both tetrahedra,
not just a simple average, implying that Eq. (\ref{aver}) requires a
correction for the internal strain that the connectedness of and interaction
between the tetrahedra produce. We refer to ref. \cite{Martin} for further
details. The resulting strain-corrected expression for the elastic moduli
tensor of the hcp lattice, $\boldsymbol{C}^{HCP}$, reads \cite{Martin} 
\begin{equation}
C_{ijkl}^{HCP}=\overline{C}_{ijkl}^{HCP}-\Delta _{ijrs}\left( \overline{C}%
_{rstu}^{HCP}\right) ^{-1}\Delta _{tukl}\;,
\end{equation}%
where $\overline{\boldsymbol{C}}^{HCP}$ is given by Eq. (\ref{aver}) and $%
\boldsymbol{\Delta }$ is a correction tensor identical to it, except that
the difference between the two components is taken instead of the sum.

There are six distinct elastic moduli in the trigonal representation of
which only three are independent in the fcc crystal and five in the hcp
crystal. The relations between the elastic moduli of the fcc lattice and
those of the hcp lattice \textit{not} corrected for the internal strain are
given by (using standard Voigt notation) 
\begin{eqnarray}
&&\ \overline{C}_{11}^{HCP}=(C_{11}^{FCC}+C_{12}^{FCC}+2C_{44}^{FCC})/2\;, 
\nonumber \\
&&\ \overline{C}_{12}^{HCP}=(C_{11}^{FCC}+5C_{12}^{FCC}-2C_{44}^{FCC})/6\;, 
\nonumber \\
&&\ \overline{C}_{13}^{HCP}=(C_{11}^{FCC}+2C_{12}^{FCC}-2C_{44}^{FCC})/3\;, 
\nonumber \\
&&\ \overline{C}_{14}^{HCP}=(C_{11}^{FCC}-C_{12}^{FCC}-2C_{44}^{FCC})/3\sqrt{%
2}\;, \\
&&\ \overline{C}_{33}^{HCP}=(C_{11}^{FCC}+2C_{12}^{FCC}+4C_{44}^{FCC})/3\;, 
\nonumber \\
&&\ \overline{C}_{44}^{HCP}=(C_{11}^{FCC}-C_{12}^{FCC}+C_{44}^{FCC})/3\;. 
\nonumber
\end{eqnarray}%
The strain-corrected constants obey 
\begin{eqnarray}
&&\ C_{11}^{HCP}=\overline{C}_{11}^{HCP}-\left( \overline{C}%
_{14}^{HCP}\right) ^{2}/\overline{C}_{44}^{HCP}\;,  \nonumber
\label{transform1} \\
&&\ C_{12}^{HCP}=\overline{C}_{12}^{HCP}+\left( \overline{C}%
_{14}^{HCP}\right) ^{2}/\overline{C}_{44}^{HCP}\;,  \nonumber \\
&&\ C_{13}^{HCP}=\overline{C}_{13}^{HCP}\;,  \nonumber \\
&&\ C_{14}^{HCP}\equiv 0\;, \\
&&\ C_{33}^{HCP}=\overline{C}_{33}^{HCP}\;,  \nonumber \\
&&\ C_{44}^{HCP}=\overline{C}_{44}^{HCP}+\left( \overline{C}%
_{14}^{HCP}\right) ^{2}/[\frac{1}{2}(\overline{C}_{11}^{HCP}-\overline{C}%
_{12}^{HCP})]\;.  \nonumber
\end{eqnarray}%
The mapping of $\boldsymbol{C}^{FCC}$ onto $\boldsymbol{C}^{HCP}$ implicit
in Eqs. (4) and (5) agrees well with experimental data on ZnS, a compound
that can crystallize both in an fcc and in an hcp lattice \cite{Martin}. In
fact, the mapping works very well for fcc and hcp crystals of hard spheres
too, as we show next.

In Figure 1, we have plotted the relative difference between the various
moduli of the fcc and hcp crystals of hard spheres as a function of the
dimensionless crystal density $\rho _{S}\sigma ^{3}$, with $\rho _{S}$\ the
number density and $\sigma $ the diameter of the spheres, using the computer
simulation data of Pronk and Frenkel (obtained from Table I of Ref. \cite%
{Pronk}) and the prediction of Martin, Eqs. (\ref{transform}--\ref%
{transform1}) \cite{Martin}. We find that the largest deviation between the
two is about 6\%. This implies that the approximate theory outlined above,
based entirely on a geometric argument, plausibly explains the difference
between the elastic moduli of fcc and hcp crystals. That is, geometry
explains almost all of the found differences in elastic behavior of fcc and
hcp crystals. 

In order to further verify Eqs. (\ref{transform}--\ref{transform1}), we
calculated the elastic moduli of the hcp hard-sphere crystal, using results
for the elastic moduli of fcc crystals of hard spheres obtained from density
functional theory (DFT) \cite{Laird}. We used the predictions of the
modified weighted-density approximation DFT, MWDA DFT, because they are
known to agree very well with the results of computer simulations. The
results of the mapping of the fcc moduli onto the hcp moduli are presented
in Figure 2, again as a function of the dimensionless density of the
spheres. The agreement with the results of the computer simulations of Pronk
and Frenkel is quite good for all moduli except $C_{13}$, for which it is
not as impressive but still satisfactory.

We believe to have demonstrated that the large difference in elasticity
between the fcc and hcp crystals of hard spheres found in \cite{Pronk} is
largely caused by the geometrical differences of these two types of crystal
lattice. We point out that even when the thermodynamic properties of fcc and
hcp crystals are similar to the point of being virtually indistinguishable,
there is in fact no reason for their elastic properties to be similar too.
The reason is that similarities in the free energy landscape at long
wavelengths do not preclude differences at short wavelengths. Indeed,
(direct) correlations at short wavelengths contribute much more
significantly to the elastic constants than to the equilibrium free energy 
\cite{Sushko}.

\bigskip \newpage

\begin{figure}[tbp]
\includegraphics[width=10cm]{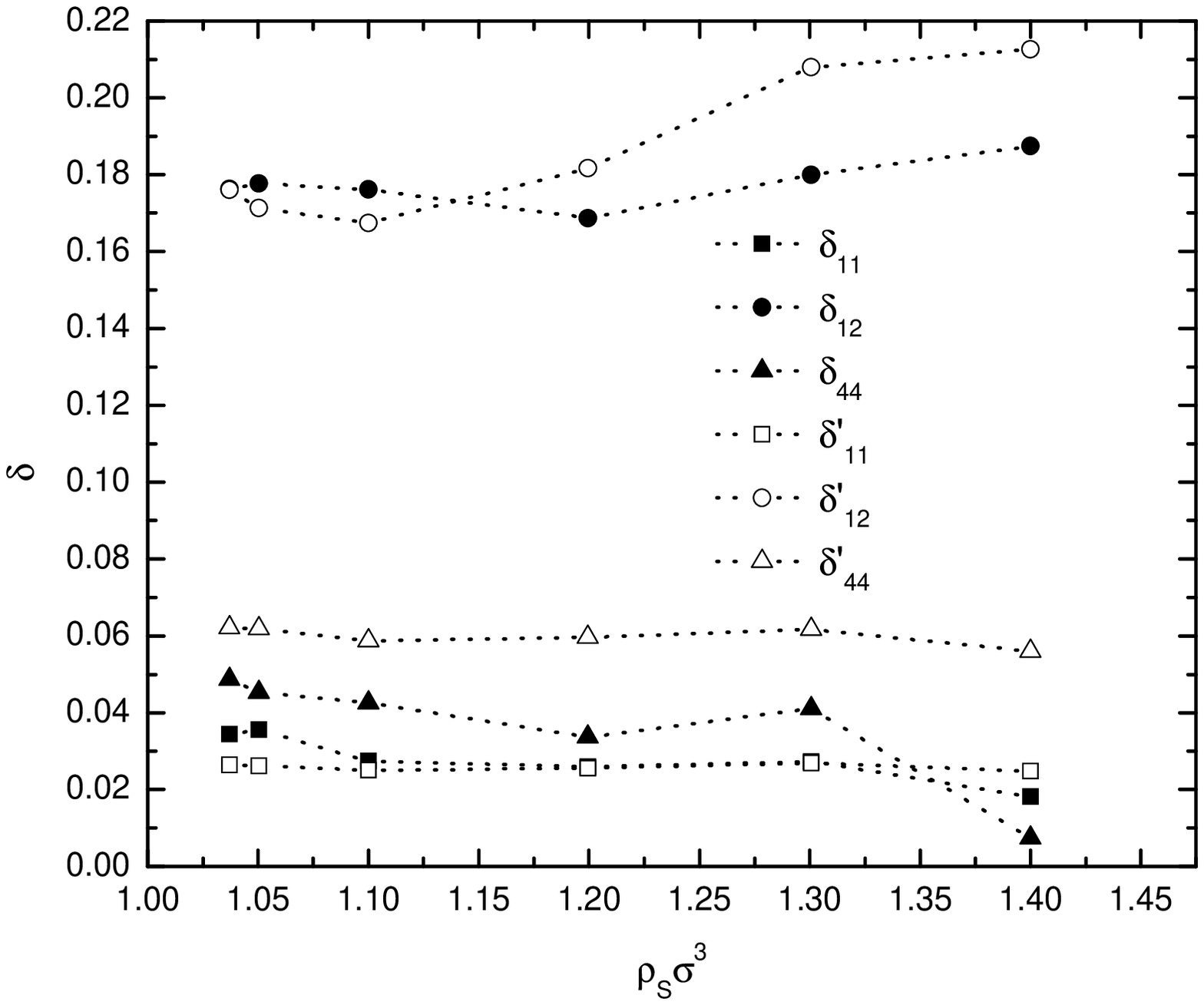}% Here is how to import EPS art
\caption{Relative difference $\protect\delta _{ij}\equiv
|C_{ij}^{FCC}-C_{ij}^{HCP}|/C_{ij}^{FCC}$ between the elastic moduli of the
hcp and fcc crystals, as a function of the dimensionless density $\protect%
\rho _{S}\protect\sigma ^{3}$. Shown are the results of computer simulations
of Ref. \protect\cite{Pronk}, $\protect\delta _{ij}$, and the ones computed
using relations Eqs. (\protect\ref{transform}--\protect\ref{transform1}), $%
\protect\delta _{ij}^{^{\prime }}$ .}
\end{figure}

\newpage

\begin{figure}[tbp]
\includegraphics[width=10cm]{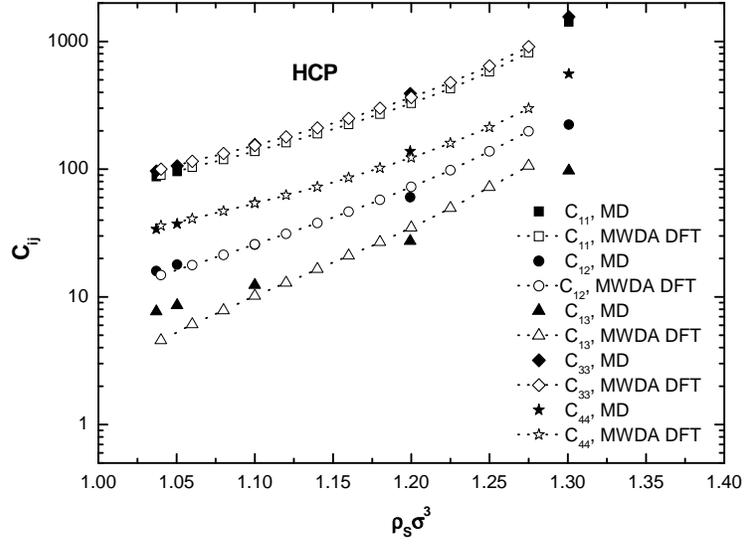}% Here is how to import EPS art
\caption{The dimensionless elastic moduli $C_{ij}\equiv C_{ij}\protect\sigma %
^{3}/k_{B}T$ of the hcp crystal computed using the relations Eqs. (\protect
\ref{transform}--\protect\ref{transform1}) and the results of the MWDA DFT
calculations for the fcc crystal \protect\cite{Laird}, as a function of the
dimensionless crystal density $\protect\rho _{S}\protect\sigma ^{3}$. Also
shown are the results of the computer simulations of Ref. \protect\cite%
{Pronk}. }
\end{figure}

\end{document}